\documentclass[12pt,aps,floats,amsmath,showpacs]{revtex4}
\input BoxedEPS
\SetRokickiEPSFSpecial  
\HideDisplacementBoxes

\newcommand{\Cite}[1]{[{\tt #1}]\cite{#1}\relax}
\let\Cite\cite

\begin{document}
 \title{Effect of revised \protect\boldmath $R_n$ measurements on
extended Gari-Kr\"umpelmann model fits  to nucleon  electromagnetic form
factors}


\author{Earle L. Lomon}
\affiliation{Center for Theoretical Physics\\
Laboratory for Nuclear Science and 
Department of Physics\\ Massachusetts Institute of Technology, Cambridge, 
Massachusetts 02139\\
\rm MIT-CTP-3765}

\begin{abstract}\noindent
The extended Gari-Kr\"umpelmann (GK) model of nucleon electromagnetic form factors, in which
the $\rho$, $\rho'$, $\omega$, $\omega'$ and  $\phi$ vector meson pole contributions evolve at  high momentum
transfer to conform to the predictions of perturbative QCD (pQCD),  was recently shown to
 provide a very good overall fit to all the nucleon electromagnetic form
 factor (emff) data, including the preliminary $R_p$ and $R_n$ polarization data available in 2002, but excluding the older $G_{Ep}$ and $G_{En}$ differential cross section data that was inconsistent with the $R_p$ and $R_n$ data. The recently published final version of the polarization data of the electric to magnetic ratio 
$R_p$  differs little from the preliminary values for the former, but the new values of $R_n$ are midway  between the preliminary values and those inferred from the differential cross section data and the Rosenbluth separation.
A new fit of the parameters of the same model has  been made with the final $R_p$ and $R_n$ data replacing the preliminary values and
 the addition of some new $R_n$ and $G_{Mn}$ data.  Again there is a good fit to all the data when excluding the differential
 cross section data that is inconsistent with the polarization data.  This includes a very good fit of the $R_p$ data, which was not
 possible when the differential cross section $G_(En)$ data was used in place of the polarization $R_n$ data.  Thus the change between
 the preliminary and final $R_n$ data, while substantial, has not impeded the good 
 simultaneous fit to the neutron and proton data.  The parameters, fit  to the data and predictions of the new model are compared to
 those of the previous models.  Low momentum structures that appear in some data analyses are partially reproduced by the model.
\end{abstract}


\maketitle



\section{INTRODUCTION}

	The Gari-Kr\"umpelmann (GK) model of nucleon electromagnetic form factors \Cite{GK} incorporates the physical features of
 meson exchange (in the vector dominance approximation) at low momentum transfers, and asymptotic freedom at high momentum 
transfers on the QCD scale.  In previous papers \Cite{EL1} and \Cite{EL2} the GK model was extended in its parameterization of the
 hadronic form factors, by including the width of the $\rho$ meson, and adding $\rho'$ and $\omega'$ exchange.  When the polarization data
 for the ratios $R_p$ and $R_n$ became available it was shown \Cite{EL2} that the GKex(02) model (Eqs. 5, 6 and 7 of \Cite{EL2}) provided
 a particularly good fit to the neutron and proton electromagnetic form factor data when polarization values for $G_{Ep}$ and $G_{En}$
 were used, replacing the inconsistent values obtained from the differential cross sections. 
 
 	 It is shown here that the quality of the fit is maintained, and the fitted parameters are closer to the physical values,
  when the revised final vesions of
 the $R_p$ and $R_n$ data, and some new data, are used.
 It is also noted that relatively narrow structures appearing in some data near $Q^2 \approx 0.2 (GeV/c^2)$ are partially reproduced by the data.
 The model parameters were negligibly influenced by the small weight of the data in the region of the structures.  
 Instead the effect can be traced to the influence of the full momentum range of the data on the sign variation of the  residues of the
 vector meson poles.

\section{The Nucleon emff model}\label{s:2}

	In fitting the nucleon emff data including the newly available $R_n$ \Cite{Madey} and $R_p$ \Cite{Perd}
results we use the extended GK model GK(02) of Ref. \cite{EL2}, as defined by Eqs. 5, 6 and 7 of \Cite{EL2}, which has the following features:
 
\begin{enumerate}
\item[(1)]  The photon-nucleon interaction is mediated by the exchange of the vector-mesons $\rho$,  $\omega$, $\phi$,
$\rho'$(1.45 GeV) and $\omega'$(1.419 GeV).
 \item[(2)]  The vector mesons are treated in the zero-width (pole) approximation, with the  exception of the $\rho$ meson.
 \item[(3)]  An approximation to the dispersion integral, that incorporates the width, is  used for the $\rho$ meson. 
 \item[(4)]  It uses the QCD cut-off $\Lambda_2$ for the helicity flip
meson-nucleon form factors, rather than the meson-nucleon cut-off $\Lambda_1$ used in previous versions.
 \item[(5)]  The evolution of the logarithmic dependence on $Q^2$ is controlled by
the quark-nucleon cut-off $\Lambda_D$, along with
$\Lambda_{\mathrm{QCD}}$.  The latter is fixed  in its experimental range at 0.15 GeV.
 \item[(6)]  Asymptotically the form factors have the  pQCD behavior (see Eq. 8 of \Cite{EL2}).
 \item[(7)] The form factor $F^\phi_1(Q^2)$ vanishes at  $Q^2 =0$, and it and also 
 $F^\phi_2(Q^2)$ decreases more rapidly at large  $Q^2$ than the other meson form factors.  This
 conforms to the Zweig rule imposed by the $s\bar s$ structure of the $\phi$ meson \cite{GK}.
\end{enumerate}
	
\section{Data base and fitting procedure}\label{s:3}

	In the fit GKex(02S) of \Cite{EL2} the data of Refs.[7-14, 16-36] in that paper was used, with the omission
of the $G_{Ep}$ values for $Q^2 \geq 1.75$ ~GeV$^2/c^2$ of that Ref. [7] and the $G_{En}$ values for 
$Q^2 \geq 0.779$ ~GeV$^2/c^2$ of Refs.[9, 17, 18] there.  In the present fit we substitute the $R_p$ values of
\Cite{Perd} and the $R_n$ values of \Cite{Madey} for the preliminary values of the Ref[5] and Ref[14] of the previous paper \Cite{EL2}. 
  As noted above, the $R_p$ values change little (although there is one more data point), while there is a
 substantial change in the $R_n$ values.  In addition we add the newly available $R_n$ data of \Cite{Warren} and
$G_{Mn}$ data of \Cite{Kubon}.

	As in the case GKex(02S) of \Cite{EL2} the 8 photon coupling constants of the four vector mesons treated as poles and 4 scale
 parameters were varied.  However the present fit was made with fixed N=1 (implying negligible error in the $\rho$ meson dispersion relation evaluation) and with
$\Lambda_{\mathrm{QCD}}$ fixed at the physical value of 0.15 ~GeV$/c$.  The 12 free parameters were optimized using a Mathematica
 program that incorporates the Levenberg-Marquardt  method.
 
	There are two  pairs of parameters which are so strongly correlated that the fit varies negligibly over large 
 ranges of the parameters when both vary in a correlated manner. In fits to the data the photon-$\rho'$ coupling,
 $g_{ \rho'}/f_{\rho'}$, is small.  The contribution of the $\rho'$ to $F_1$ is negligible, but if $\kappa_{\rho'}$ is
large the $\rho'$ contribution to $F_2$ is of importance and best fits the data when the product of the two parameters is
near .08 .  For $\kappa_{\rho'}$ running from 12.0 to infinity, $g_{\rho'}/f_{\rho'}$ runs from .007 to zero.  For this
 whole range the $\chi^2$ value only changes by $\leq 0.1$.  We constrain $\kappa_{\rho'}$ to the smallest value in this range with the correlated largest value of
  $g_{\rho'}/f_{\rho'}$.
 
 	 The other strongly correlated parameter pair is $\kappa_\phi$ and $\mu_\phi$ which appear together in the $\phi$ hadronic
form factor $F_2^\phi$.  In the fits $\mu_\phi$ is small ($\leq 0.2GeV/c$).  Consequently for Q$\geq 0.5 GeV/c$, $F_2^\phi$ is
 approximately proportional to $\kappa_\phi$/$\mu_\phi$.  As $\mu_\phi$ runs from 0.2 to zero, $\kappa_\phi$ runs from
0.01 to zero and the best fit $\chi^2$ again only changes $\leq 0.1$.  Constraining both $\mu_\phi$ to 0.2 and $\kappa_\phi$ to
 0.01 only increases the  best fit $\chi^2$ by 0.06.  Only the remaining 8 parameters are individually sensitive to the data fit.
 
 	 The resulting parameters and the quality of the fit to this data set, GKex(05), is
compared with that of the GKex(02S) and GKex(01) models in Tables I and II. The fit of these models to the complete data set
(including the differential cross section data for electric form factors inconsistent with polarization data) is 
 displayed in Figs. 1-6. The model  GKex(01), which did not include the $\omega'$ meson, was fitted to all the differential cross section data, 
including the available $R_p$ data but in the absence of the $R_n$ data.  The present GKex(05) model differs from GKex(02S) only due to the
substitution of the newer polarization data and inclusion of the few new $R_n$ and $G_{Mn}$ points.

\begin{table}[p]
 \caption{Model parameters.  Common to all models are $\kappa_v=3.706$,
$\kappa_s=-0.12$, $m_\rho=0.776$ GeV, $m_\omega=0.784$ GeV, $m_\phi=1.019$
GeV,  $m_{\rho'}= 1.45$ GeV and $m_{\omega'}= 1.419$ GeV.}
 \label{T1}
 \begin{tabular}{c|cccc}
Parameters & & Models &  \\
\hline\hline
 &  GKex(01)&  GKex(02S)  &  GK(05)  \\
$g_{\rho'}/f_{\rho'}$ & 0.0636 & 0.0401 & 0.007208 \\
$\kappa_{\rho'}$ & $-0.4175$ & 6.8190 & 12.0 \\
$g_\omega/f_\omega$ & 0.7918 & 0.6739 & 0.7021 \\
$\kappa_\omega$ & 5.1109 & 0.8762 & 0.4027 \\
$g_\phi/f_\phi$ & $-0.3011$ & $-0.1676$ & $-0.1711$ \\  
$\kappa_\phi$ & 13.4385 &  7.0172 & 0.01 \\ 
$\mu_\phi$ & 1.1915 & 0.8544 & 0.2 \\
$g_{\omega'}/f_{\omega'}$ &  & 0.2552 & 0.164 \\
$\kappa_{\omega'}$ &  & 1.4916 &  $-2.973$ \\
$\Lambda_1$ & 0.9660  & 0.9407 & 0.93088 \\ 
$\Lambda_D$ & 1.3406  & 1.2111 &1.181 \\ 
$\Lambda_2$ & 2.1382 & 2.7891 & 2.6115 \\
$\Lambda_{\mathrm{QCD}}$ & 0.1163 & 0.150 (a) & 0.150 (a) \\
N & 1.0 (a) & 1.0 (a) & 1.0 (a)
 \end{tabular} \\
 (a) not varied
 \end{table}%

\begin{table}[p] 
 \caption{Contributions to the standard deviation, $\chi^2$ for each form factor, from the present data set (Section 3) for
 each of the three models. The number of data points contributing is in the second column.  The polarization data is only used for the
$R_{n,p}$ values and only differential cross section data is used for the $G_{Mp.Ep,Mn,En}$ values (and cold neutron scattering for the
$G_{En}$slope at Q=0).}
 \label{T2}
 \begin{tabular}{r|c|rrrr} 
Data & Data & & Models &  \\[-0.5ex]
type & size & GKex(01) & GKex(02S) & GKex(05) \\
\hline\hline 
$G_{Mp}$ & 68 & 43.3 & 47.9 & 51.5 \\
$G_{Ep}$ & 36 & 26.9 & 30.5 & 28.4 \\
$G_{Mn}$ & 39 & 127.8 & 129.4 & 124.9 \\
$G_{En}$ & 15 & 25.3 & 24.1 & 25.1 \\
$R_p$ & 22 & 135.8 & 10.4 & 10.3 \\
$R_n$ & 5 & 1.9 & 5.6 & 1.1 \\
\hline 
Total & 185 & 361.0 & 248.0 & 241.3 \\
\end{tabular}
\end{table} 

\section{Results}\label{s:4}
	Table~\ref{T1} presents the parameters  which minimize  $\chi^2$ for the above 3 cases. to their own data sets  For all 3 parameter sets the  hadronic form factor cut-off masses, $\Lambda_1$, 
$\Lambda_2$, $\Lambda_D$,  and $\mu_\phi$. are reasonable.  The relatively large value of
$\Lambda_2$, which controls the spin-flip suppression in QCD, is consistent with the slow approach
 to asymptopia observed in polarized hadron scattering.  For the case in which $\Lambda_{QCD}$
is a fitted parameter, as well as the two for which it is fixed, it is consistent with high energy experiment.  The addition of the $\omega'$(1.419) meson in GKex(02S) and especially in GKex(05) has
 moved $\kappa_\omega$ closer to the
 expected small negative value than all earlier fits, but the remaining small positive value implies some
 contribution may be required
 from a still higher mass isoscalar meson.  However the adequacy of the fits is an indication that form factors
 with more poles would produce similar fits to those already obtained.

	In Table~\ref{T2} the values of $\chi^2$ for the present data set are listed for the three cases, detailing the 
 contribution from each of the six form factors.  The $\chi^2$ values for $G_{Mp},$ $G_{Ep}$, $R_p$ and $R_n$ are less than the number of data points, but for $G_{Mn}$, and to  a lesser extent for $G_{En}$, the $\chi^2$ values are substantially larger
 than the number of data points.  These large values of $\chi^2$ are caused by the inconsistent fluctuations of the data from
 point to point for $Q^2 \leq 2$ ~GeV$^2/c^2$. Fitting the data better in that range would require unreasonable oscillations in a model. 

 \begin{figure}[hbt]
$$
\BoxedEPSF{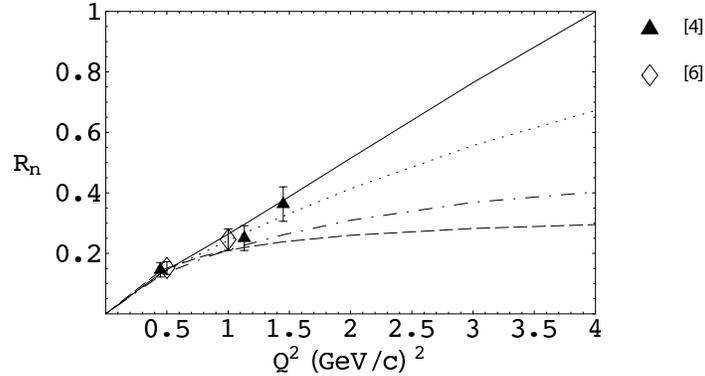 }
$$
\caption{$R_n$, the ratio $\mu_n G_{En}/G_{Mn}$.  Comparison of the models 
GKex(01) [solid], GKex(02S) [dash-dotted] and GKex(05) [dotted] with the data. 
The dashed curve is  $R^{Galster}_n(Q^2)$ of Eq. 1 of \protect\Cite{EL2}.  The experimental points are
described in the text \protect\Cite{Madey},\protect\Cite{Warren}.  The data symbols are listed beside the  figure.}
\label{elFig1}
\end{figure}

 Fig. 1 shows the substantial effect on the best fit model of the new highest momentum transfer polarization $R_n$ value \Cite{Madey}.  The experimental
 values and the model predictions are substantially larger than the phenomenological Galster prediction, but far less than the values 
determined from the differential cross section data. The validity of the model up to 3.5 $ ~GeV ^2/c^2$ will be tested when recent JLab data \Cite{E02013} has been analyzed. 
 However, as seen in Fig. 2, the substantial change in the model parameters required to fit the new $R_n$ data (together with new $R_p$ and  $G_{Mn}$ data) has only a small effect on the fit to the $R_p$ data.  This is  remarkable when compared to the much poorer fit to the $R_p$
 data required by fitting the GKex(01) data set, which differs from the $R_n$ GKex(01) curve by about the same amount as the latter differs from the GKex(02S) curve.  As shown in Figs. 3-6 there is little change in the  $G_{Ep}$ and $G_{Mp}$ predictions up to large momentum transfers, while $G_{Mn}$ and $G_{En}$ are substantially altered at the larger momentum transfers.

 \begin{figure}[hbt]
$$
\BoxedEPSF{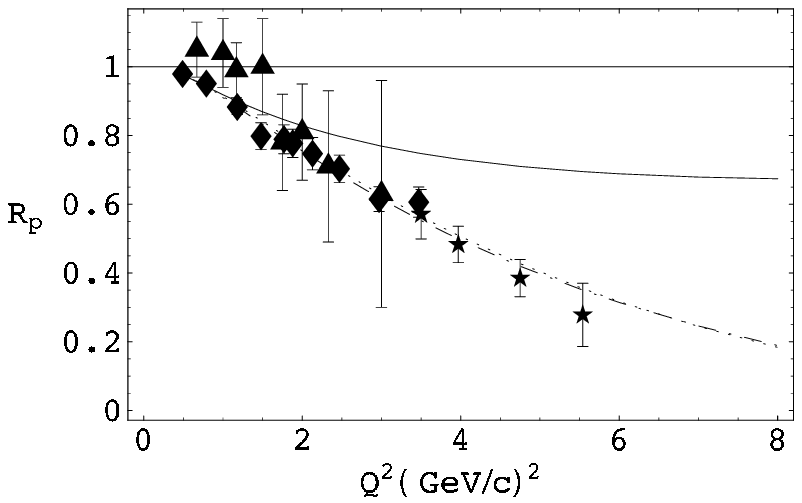 scaled 900} \qquad \BoxedEPSF{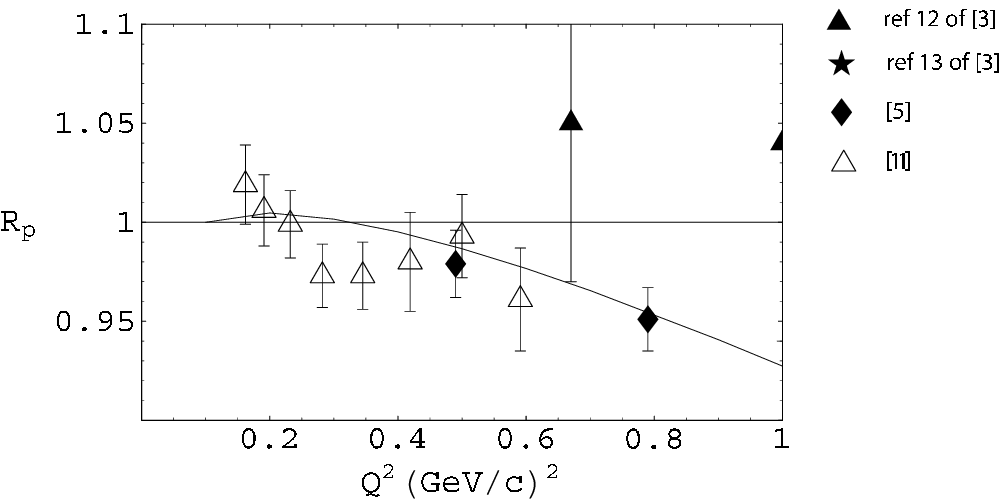 scaled 900}
$$
\text{\bf (a)} \hspace{3in} \text{\bf (b)}
\caption{$R_p$, the ratio $\mu_p G_{Ep}/G_{Mp}$.  (a) Comparison of the models
GKex(01) [solid], GKex(02S) [dash-dotted] and GKex(05) [dotted] with the data.
The polarization data is from  \protect\Cite{Perd}, and from Ref.[12] of \protect\Cite{EL2}.   (b) Comparison of model GKex(05) [dotted] with nth new data of  \protect\Cite{Crawford}.  The data symbols are listed beside the  figure.}
\label{elFig2}
\end{figure}

 \begin{figure}[hbt] 
$$
\BoxedEPSF{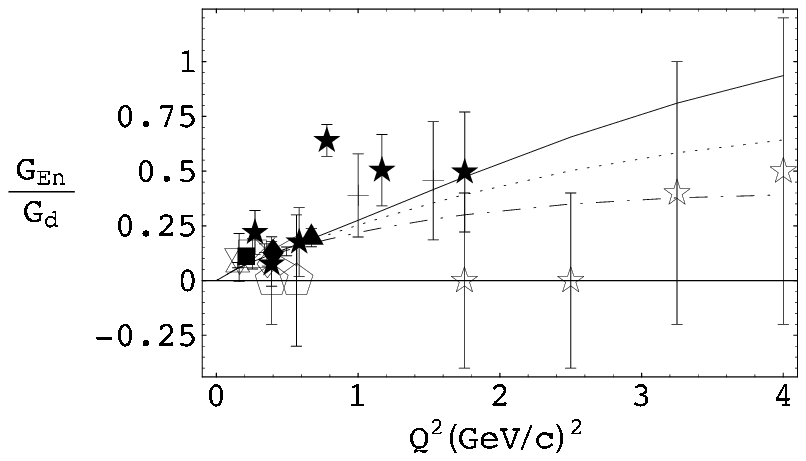 scaled 900} \qquad \BoxedEPSF{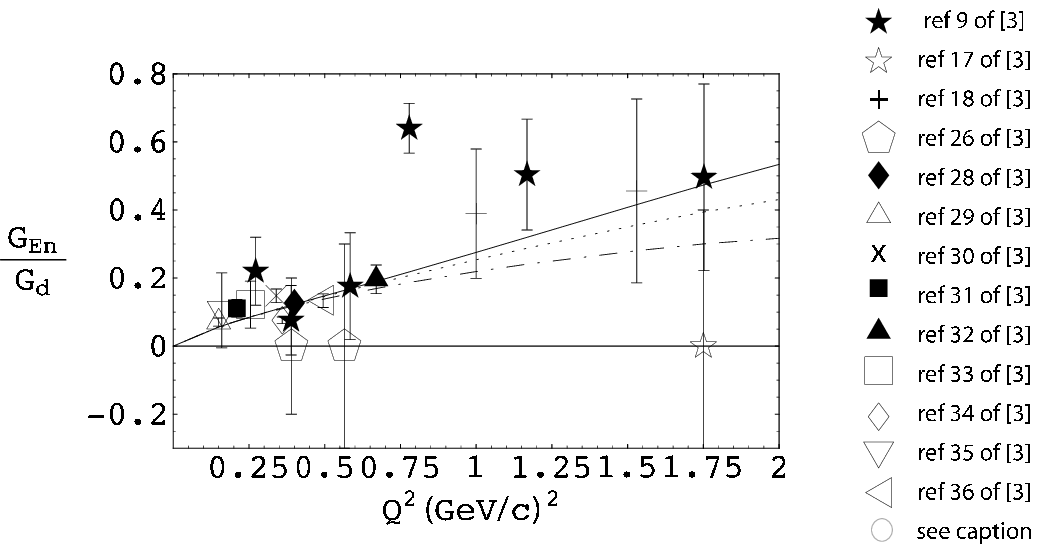 scaled 900}
$$
\text{\bf (a)} \hspace{3in} \text{\bf (b)}
\caption{i(a) $G_{En}$ normalized to $G_{d}$.  Comparison of the models 
GKex(01) [solid], GKex(02S) [dash-dotted] and GKex(05) [dotted] with the data  of Refs. [9, 17, 18. 26, \& 28-36] of 
\protect\Cite{EL2}.  The data symbols are listed beside the  figure.  The points labelled by open
circles are obtained by multiplying $R_n$ data \protect\Cite{Madey} and \protect\Cite{Warren} by the $G_{Mn}$ of GKex(05)
normalized by $\mu_n G_d$
  (b) Expansion of the range to $Q^2 \leq2.0$~GeV$^2/c^2$.}
\label{elFig3}
\end{figure}

 \begin{figure}[hbt]
$$
\BoxedEPSF{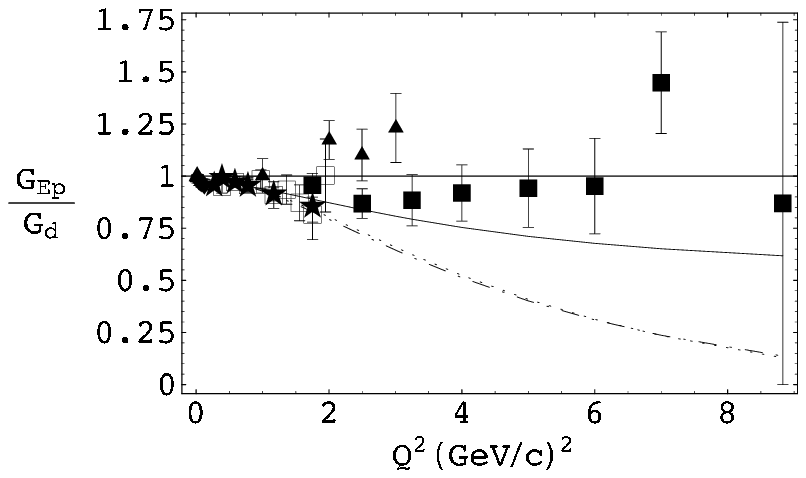 scaled 900} \qquad \BoxedEPSF{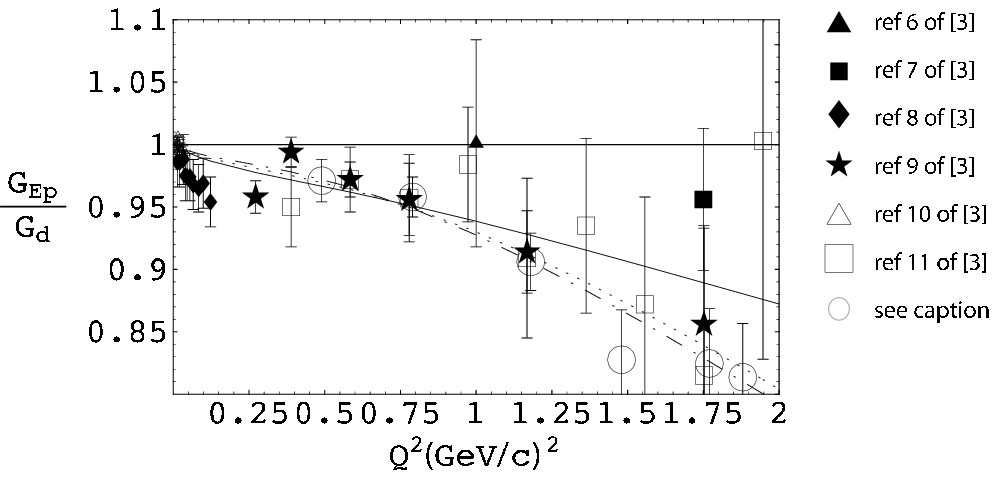 scaled 900}
$$
\text{\bf (a)} \hspace{3in} \text{\bf (b)}
\caption{$G_{Ep}$ normalized to $G_d$.  Comparison of the models 
GKex(01) [solid], GKex(02S) [dash-dotted] and GKex(05) [dotted] with the data of Refs. [6-11] of \protect\Cite{EL2}. 
 The data symbols are listed beside the  figure.
The points labelled by open
circles are obtained by multiplying $R_p$ data of \protect\Cite{Perd} and Ref. 13 of \protect\Cite{EL2} by the $G_{Mp}$ of GKex(05)
normalized by $\mu_p G_d$
 (a) The full data range.  (b) Expansion of the range
 $Q^2 \leq 2.0$~GeV$^2/c^2$.}
\label{elFig4}
\end{figure}

\begin{figure}[hbt]
$$
\BoxedEPSF{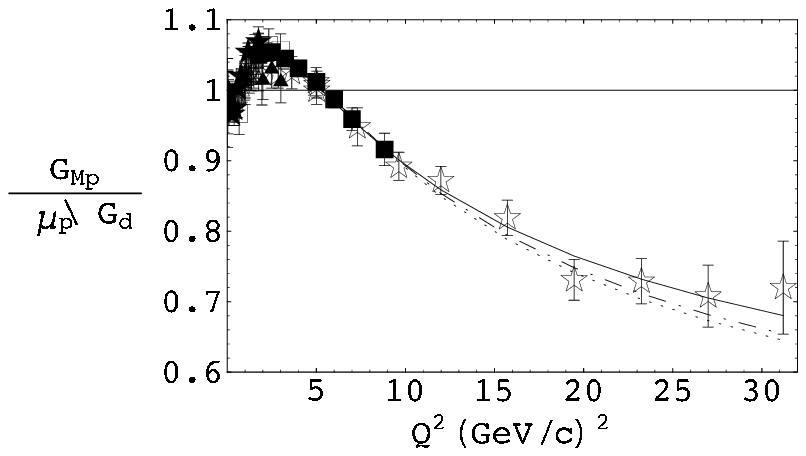 scaled 900} \qquad \BoxedEPSF{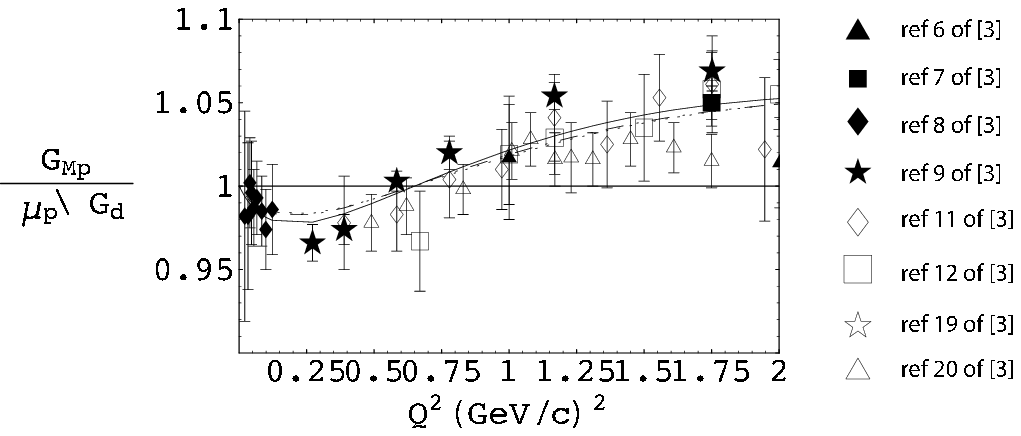 scaled 900}
$$
\text{\bf (a)} \hspace{3in} \text{\bf (b)}
\caption{$G_{Mp}$ normalized to $\mu_p G_d$.  Comparison of the models 
GKex(01) [solid], GKex(02S) [dash-dotted] and GKex(05) [dotted] with the data of Refs [6-9, 11, 12, 19 \& 20] of \protect\Cite{EL2},
.  The data symbols are listed beside the  figure.  (a) The full data range.  (b) Expansion of the range  
$Q^2 \leq 2.0$~GeV$^2/c^2$.}
\label{elFig5}
\end{figure}

\begin{figure}[hbt] 
$$
\BoxedEPSF{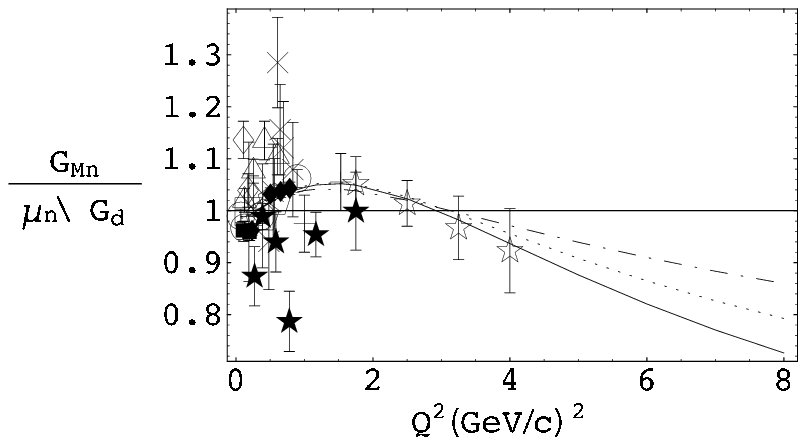 scaled 900}\qquad \BoxedEPSF{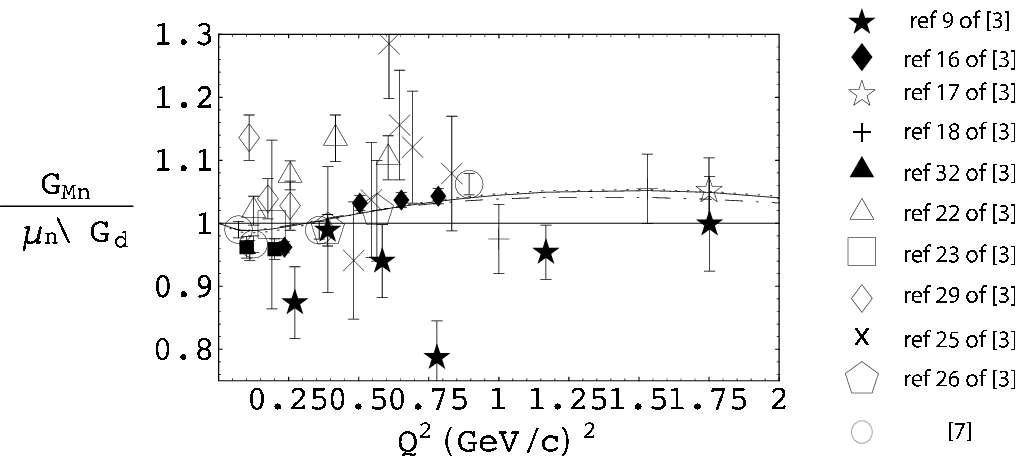 scaled 900}
$$
\text{\bf (a)} \hspace{3in} \text{\bf (b)}
\caption{$G_{Mn}$ normalized to $\mu_n G_{d}$.  Comparison of the models 
GKex(01) [solid], GKex(02S) [dash-dotted] and GKex(05) [dotted] with the data of \protect\Cite{Kubon} 
 and of Refs. [9, 16-18, \& 21-27] of \protect\Cite{EL2}.
The data symbols are listed beside the figure.  
 (a) The full data range.  (b) Expansion of the range  
$Q^2 \leq 2.0$~GeV$^2/c^2$.}
\label{elFig6}
\end{figure}

	It is noteworthy that, while the parameters of this model were fitted to the whole momentum transfer region of the available data, 
the model reproduces, at least in part, narrow structure observed at very low momentum transfer, between 0.1 and 0.4 $~GeV ^2/c^2$.
 This was noted in the data available in 2003 by \Cite{F&W}, and may be confirmed by recent data taken with the BLAST detector at 
Bates Laboratory \Cite {Milner}.  In \Cite {F&W} the structure is phenomenologically related to a pion cloud. In a VDM model, such as 
the ones discussed here,  the pion cloud is represented by pion pairs and triplets clustered into vector mesons.  Figs. 3-6 (b) show 
the data in the expanded scale of $Q^2 \leq 2.0$~GeV$^2/c^2$ to better observe the small structures in the data and models.  For both 
$G_{Mp}$ in Fig. 5(b) and $G_{Mn}$ in Fig. 6(b) a minimum appears near $Q^2 \approx 0.2$~GeV$^2/c^2$ in the published data and is 
reproduced by the models of this article.  For $G_{Ep}$ the data also shows a minimum in the same vicinity, but our models, Fig. 4(b), 
only have a horizontal inflection there.  Our models show no low momentum-transfer structures for $G_{En}$, Fig. 3(b), and the presence of the
  maximum indicated in \Cite {F&W} is not clear in the published data.  The recent experiments \Cite {Milner} may provide a 
  clarification.  The proton polarization data has now become available \Cite{Crawford}, too late to be
   included in the model parameter fitting.  That data is compared with model GKex(05) in Fig. 2(b), showing two data points (out of eight) that are lower than the model curve by about i.5 standard deviations.  To the extent that the model does have structure near $Q^2 \approx 0.2$~GeV$^2/c^2$ the
    effect can be traced to the opposite signs of the $\omega$ and $\omega'$ meson contributions to $F_2^{is}$ combined with the more rapid decrease of the $\omega$ contribution in that region.

\section{Conclusions}\label{s:5}

The incorporation of data that has become available since the publication of \Cite {EL2}, especially the $R_n$ data of \Cite {Madey}, 
was used  to refit the parameters GKex(02S), resulting in the version GKex(05).  Only the predictions for $R_n$ and $G_{En}$ at higher
 momentum-transfers were substantially changed, and a good fit to the data maintained.  The predicted values of $R_n$ continue to rise
 above the Galster curve with increasing momentum transfer, a prediction which may soon be checked by new data \Cite {E02013}.
 The fitted parameters agree with known 
 constraints and the model is consistent with VMD at lower momentum-transfers, while approaching pQCD at high momentum-transfers. 
  
	It also has been noted that the predictions of the model include at least some of the structure  recently noted near  $Q^2 \approx 0.2$~GeV$^2/c^2$.

\subsection*{Acknowledgments}
	The author is grateful to Haiyan Gao, Richard Milner, Christopher Crawford, Jason Seely and Will Brooks for being  kept informed of 
recent experiments.

This work is supported in part by funds provided by the U.S. Department of Energy (DOE) under cooperative research agreement DE-FC02-94ER40818.

\end{document}